\documentclass[sigconf,letterpaper,10pt,dvipsnames]{acmart}
\usepackage[T1]{fontenc}
\usepackage[utf8]{inputenc}
\usepackage{microtype}

\usepackage{amsmath}
\usepackage{mathtools}

\usepackage{enumitem}
\usepackage{xspace}
\usepackage{balance}
\usepackage{soul}
\usepackage{listings}
\usepackage{xcolor}

\definecolor{codegray}{rgb}{0.25,0.25,0.25} 
\definecolor{codepurple}{rgb}{0.58,0,0.82}

\lstdefinestyle{mystyle-mgen}{
  commentstyle=\color{gray},
  keywordstyle=\color{MidnightBlue},
  numberstyle=\tiny\color{codegray},
  stringstyle=\color{codepurple},
  basicstyle=\color{Periwinkle}\ttfamily\scriptsize,
  rulecolor=\color{black},
  breakatwhitespace=true,         
  breaklines=true,                 
  captionpos=b,
  frame=tb,
  keepspaces=true,                 
  numbers=left,                    
  numbersep=5pt,                  
  showspaces=false,                
  showstringspaces=false,
  showtabs=false,                  
  tabsize=2,
  xleftmargin=10pt,
  belowskip=-10pt,
}

\lstdefinelanguage{mgen}{
  alsoletter={-},
  keywords={listen,on,off,dst,periodic},
  sensitive=false,
  comment=[l]{\#},
  morecomment=[s]{/*}{*/},
  moredelim=[l][\color{orange}]{\&},
  moredelim=[l][\color{magenta}]{*},
  moredelim=**[il][\color{purple}{:}\color{MidnightBlue}]{:},   
  morestring=[b]',
  morestring=[b]",
}

\usepackage[acronyms,nonumberlist,nopostdot,nomain,nogroupskip,acronymlists={hidden}]{glossaries}
\newglossary[algh]{hidden}{acrh}{acnh}{Hidden Acronyms}
\glsdisablehyper

\usepackage{booktabs}
\usepackage{tabularx}
\usepackage{multirow}

\newcommand{\openrangym}{OpenRAN Gym\xspace}

\newcommand{\bs}{\gls{bs}\xspace}

\newcommand{\datasethours}{500 hours\xspace}
\newcommand{\datasetsize}{$450$\:GB\xspace}
\newcommand{\datasetkpm}{35\xspace}

\usepackage{tikz}
\usepackage{pgfplots}
\pgfplotsset{compat=newest}
\pgfplotsset{plot coordinates/math parser=false}
\newlength\fheight
\newlength\fwidth
\usetikzlibrary{plotmarks,patterns,decorations.pathreplacing,backgrounds,calc,arrows,arrows.meta,spy,matrix,scopes,patterns.meta}
\usepgfplotslibrary{patchplots,groupplots}
\usepackage{tikzscale}

\newif\ifexttikz
\exttikzfalse

\ifexttikz
	\usetikzlibrary{external}
\fi

\usepackage[font=footnotesize]{subcaption}
\usepackage[font=footnotesize]{caption}

\usepackage{xurl}

\usepackage{orcidlink}

\newacronym{3gpp}{3GPP}{3rd Generation Partnership Project}
\newacronym{4g}{4G}{4th generation}
\newacronym{5g}{5G}{5th generation}
\newacronym{6g}{6G}{6th generation}
\newacronym{5gc}{5GC}{5G Core}
\newacronym{adc}{ADC}{Analog to Digital Converter}
\newacronym{aerpaw}{AERPAW}{Aerial Experimentation and Research Platform for Advanced Wireless}
\newacronym{ai}{AI}{Artificial Intelligence}
\newacronym{aimd}{AIMD}{Additive Increase Multiplicative Decrease}
\newacronym{am}{AM}{Acknowledged Mode}
\newacronym{amc}{AMC}{Adaptive Modulation and Coding}
\newacronym{amf}{AMF}{Access and Mobility Management Function}
\newacronym{aops}{AOPS}{Adaptive Order Prediction Scheduling}
\newacronym{api}{API}{Application Programming Interface}
\newacronym{apn}{APN}{Access Point Name}
\newacronym{ap}{AP}{Application Protocol}
\newacronym{aqm}{AQM}{Active Queue Management}
\newacronym{ausf}{AUSF}{Authentication Server Function}
\newacronym{avc}{AVC}{Advanced Video Coding}
\newacronym{awgn}{AGWN}{Additive White Gaussian Noise}
\newacronym{balia}{BALIA}{Balanced Link Adaptation Algorithm}
\newacronym{bbu}{BBU}{Base Band Unit}
\newacronym{bdp}{BDP}{Bandwidth-Delay Product}
\newacronym{ber}{BER}{Bit Error Rate}
\newacronym{bf}{BF}{Beamforming}
\newacronym{bler}{BLER}{Block Error Rate}
\newacronym{brr}{BRR}{Bayesian Ridge Regressor}
\newacronym{bs}{BS}{Base Station}
\newacronym{bsr}{BSR}{Buffer Status Report}
\newacronym{bss}{BSS}{Business Support System}
\newacronym{ca}{CA}{Carrier Aggregation}
\newacronym{caas}{CaaS}{Connectivity-as-a-Service}
\newacronym{cb}{CB}{Code Block}
\newacronym{cc}{CC}{Congestion Control}
\newacronym{ccid}{CCID}{Congestion Control ID}
\newacronym{cco}{CC}{Carrier Component}
\newacronym{cd}{CD}{Continuous Delivery}
\newacronym{cdd}{CDD}{Cyclic Delay Diversity}
\newacronym{cdf}{CDF}{Cumulative Distribution Function}
\newacronym{cdn}{CDN}{Content Distribution Network}
\newacronym{cli}{CLI}{Command-line Interface}
\newacronym{cn}{CN}{Core Network}
\newacronym{codel}{CoDel}{Controlled Delay Management}
\newacronym{comac}{COMAC}{Converged Multi-Access and Core}
\newacronym{cord}{CORD}{Central Office Re-architected as a Datacenter}
\newacronym{cornet}{CORNET}{COgnitive Radio NETwork}
\newacronym{cosmos}{COSMOS}{Cloud Enhanced Open Software Defined Mobile Wireless Testbed for City-Scale Deployment}
\newacronym{cots}{COTS}{Commercial Off-the-Shelf}
\newacronym{cp}{CP}{Control Plane}
\newacronym{cyp}{CP}{Cyclic Prefix}
\newacronym{up}{UP}{User Plane}
\newacronym{cpu}{CPU}{Central Processing Unit}
\newacronym{cqi}{CQI}{Channel Quality Information}
\newacronym{cr}{CR}{Cognitive Radio}
\newacronym{cran}{CRAN}{Cloud \gls{ran}}
\newacronym{crs}{CRS}{Cell Reference Signal}
\newacronym{csi}{CSI}{Channel State Information}
\newacronym{csirs}{CSI-RS}{Channel State Information - Reference Signal}
\newacronym{cu}{CU}{Central Unit}
\newacronym{d2tcp}{D$^2$TCP}{Deadline-aware Data center TCP}
\newacronym{d3}{D$^3$}{Deadline-Driven Delivery}
\newacronym{dac}{DAC}{Digital to Analog Converter}
\newacronym{dag}{DAG}{Directed Acyclic Graph}
\newacronym{das}{DAS}{Distributed Antenna System}
\newacronym{dash}{DASH}{Dynamic Adaptive Streaming over HTTP}
\newacronym{dc}{DC}{Dual Connectivity}
\newacronym{dccp}{DCCP}{Datagram Congestion Control Protocol}
\newacronym{dce}{DCE}{Direct Code Execution}
\newacronym{dci}{DCI}{Downlink Control Information}
\newacronym{dctcp}{DCTCP}{Data Center TCP}
\newacronym{dl}{DL}{Downlink}
\newacronym{dmr}{DMR}{Deadline Miss Ratio}
\newacronym{dmrs}{DMRS}{DeModulation Reference Signal}
\newacronym{drlcc}{DRL-CC}{Deep Reinforcement Learning Congestion Control}
\newacronym{drs}{DRS}{Discovery Reference Signal}
\newacronym{du}{DU}{Distributed Unit}
\newacronym{e2e}{E2E}{end-to-end}
\newacronym{earfcn}{EARFCN}{E-UTRA Absolute Radio Frequency Channel Number}
\newacronym{ecaas}{ECaaS}{Edge-Cloud-as-a-Service}
\newacronym{ecn}{ECN}{Explicit Congestion Notification}
\newacronym{edf}{EDF}{Earliest Deadline First}
\newacronym{embb}{eMBB}{Enhanced Mobile Broadband}
\newacronym{empower}{EMPOWER}{EMpowering transatlantic PlatfOrms for advanced WirEless Research}
\newacronym{enb}{eNB}{evolved Node Base}
\newacronym{endc}{EN-DC}{E-UTRAN-\gls{nr} \gls{dc}}
\newacronym{epc}{EPC}{Evolved Packet Core}
\newacronym{eps}{EPS}{Evolved Packet System}
\newacronym{es}{ES}{Edge Server}
\newacronym{etsi}{ETSI}{European Telecommunications Standards Institute}
\newacronym[firstplural=Estimated Times of Arrival (ETAs)]{eta}{ETA}{Estimated Time of Arrival}
\newacronym{eutran}{E-UTRAN}{Evolved Universal Terrestrial Access Network}
\newacronym{faas}{FaaS}{Function-as-a-Service}
\newacronym{fapi}{FAPI}{Functional Application Platform Interface}
\newacronym{fdd}{FDD}{Frequency Division Duplexing}
\newacronym{fdm}{FDM}{Frequency Division Multiplexing}
\newacronym{fdma}{FDMA}{Frequency Division Multiple Access}
\newacronym{fed4fire}{FED4FIRE+}{Federation 4 Future Internet Research and Experimentation Plus}
\newacronym{fir}{FIR}{Finite Impulse Response}
\newacronym{fit}{FIT}{Future \acrlong{iot}}
\newacronym{fpga}{FPGA}{Field Programmable Gate Array}
\newacronym{fr2}{FR2}{Frequency Range 2}
\newacronym{fs}{FS}{Fast Switching}
\newacronym{fscc}{FSCC}{Flow Sharing Congestion Control}
\newacronym{ftp}{FTP}{File Transfer Protocol}
\newacronym{fw}{FW}{Flow Window}
\newacronym{ge}{GE}{Gaussian Elimination}
\newacronym{gnb}{gNB}{Next Generation Node Base}
\newacronym{gop}{GOP}{Group of Pictures}
\newacronym{gpr}{GPR}{Gaussian Process Regressor}
\newacronym{gpu}{GPU}{Graphics Processing Unit}
\newacronym{gtp}{GTP}{GPRS Tunneling Protocol}
\newacronym{gtpc}{GTP-C}{GPRS Tunnelling Protocol Control Plane}
\newacronym{gtpu}{GTP-U}{GPRS Tunnelling Protocol User Plane}
\newacronym{gtpv2c}{GTPv2-C}{\gls{gtp} v2 - Control}
\newacronym{gw}{GW}{Gateway}
\newacronym{harq}{HARQ}{Hybrid Automatic Repeat reQuest}
\newacronym{hetnet}{HetNet}{Heterogeneous Network}
\newacronym{hh}{HH}{Hard Handover}
\newacronym{hol}{HOL}{Head-of-Line}
\newacronym{hqf}{HQF}{Highest-quality-first}
\newacronym{hss}{HSS}{Home Subscription Server}
\newacronym{http}{HTTP}{HyperText Transfer Protocol}
\newacronym{ia}{IA}{Initial Access}
\newacronym{iab}{IAB}{Integrated Access and Backhaul}
\newacronym{ic}{IC}{Incident Command}
\newacronym{ietf}{IETF}{Internet Engineering Task Force}
\newacronym{imsi}{IMSI}{International Mobile Subscriber Identity}
\newacronym{imt}{IMT}{International Mobile Telecommunication}
\newacronym{iot}{IoT}{Internet of Things}
\newacronym{ip}{IP}{Internet Protocol}
\newacronym{itu}{ITU}{International Telecommunication Union}
\newacronym{kpi}{KPI}{Key Performance Indicator}
\newacronym{kpm}{KPM}{Key Performance Measurement}
\newacronym{kvm}{KVM}{Kernel-based Virtual Machine}
\newacronym{los}{LOS}{Line-of-Sight}
\newacronym{lsm}{LSM}{Link-to-System Mapping}
\newacronym{lstm}{LSTM}{Long Short Term Memory}
\newacronym{lte}{LTE}{Long Term Evolution}
\newacronym{lxc}{LXC}{Linux Container}
\newacronym{m2m}{M2M}{Machine to Machine}
\newacronym{mac}{MAC}{Medium Access Control}
\newacronym{manet}{MANET}{Mobile Ad Hoc Network}
\newacronym{mano}{MANO}{Management and Orchestration}
\newacronym{mc}{MC}{Multi-Connectivity}
\newacronym{mcc}{MCC}{Mobile Cloud Computing}
\newacronym{mchem}{MCHEM}{Massive Channel Emulator}
\newacronym{mcs}{MCS}{Modulation and Coding Scheme}
\newacronym{mec}{MEC}{Multi-access Edge Computing}
\newacronym{mec2}{MEC}{Mobile Edge Cloud}
\newacronym{mfc}{MFC}{Mobile Fog Computing}
\newacronym{mgen}{MGEN}{Multi-Generator}
\newacronym{mi}{MI}{Mutual Information}
\newacronym{mib}{MIB}{Master Information Block}
\newacronym{miesm}{MIESM}{Mutual Information Based Effective SINR}
\newacronym{mimo}{MIMO}{Multiple Input, Multiple Output}
\newacronym{ml}{ML}{Machine Learning}
\newacronym{mlr}{MLR}{Maximum-local-rate}
\newacronym[plural=\gls{mme}s,firstplural=Mobility Management Entities (MMEs)]{mme}{MME}{Mobility Management Entity}
\newacronym{mmtc}{mMTC}{Massive Machine-Type Communications}
\newacronym{mmwave}{mmWave}{millimeter wave}
\newacronym{mpdccp}{MP-DCCP}{Multipath Datagram Congestion Control Protocol}
\newacronym{mptcp}{MPTCP}{Multipath TCP}
\newacronym{mr}{MR}{Maximum Rate}
\newacronym{mrdc}{MR-DC}{Multi \gls{rat} \gls{dc}}
\newacronym{mse}{MSE}{Mean Square Error}
\newacronym{mss}{MSS}{Maximum Segment Size}
\newacronym{mt}{MT}{Mobile Termination}
\newacronym{mtd}{MTD}{Machine-Type Device}
\newacronym{mtu}{MTU}{Maximum Transmission Unit}
\newacronym{mumimo}{MU-MIMO}{Multi-user \gls{mimo}}
\newacronym{mvno}{MVNO}{Mobile Virtual Network Operator}
\newacronym{nalu}{NALU}{Network Abstraction Layer Unit}
\newacronym{nas}{NAS}{Network Attached Storage}
\newacronym{nat}{NAT}{Network Address Translation}
\newacronym{nbiot}{NB-IoT}{Narrow Band IoT}
\newacronym{nfv}{NFV}{Network Function Virtualization}
\newacronym{nfvi}{NFVI}{Network Function Virtualization Infrastructure}
\newacronym{ni}{NI}{Network Interfaces}
\newacronym{nic}{NIC}{Network Interface Card}
\newacronym{nlos}{NLOS}{Non-Line-of-Sight}
\newacronym{now}{NOW}{Non Overlapping Window}
\newacronym{nsm}{NSM}{Network Service Mesh}
\newacronym[type=hidden]{nr}{NR}{New Radio}
\newacronym{nrf}{NRF}{Network Repository Function}
\newacronym{nsa}{NSA}{Non Stand Alone}
\newacronym{nse}{NSE}{Network Slicing Engine}
\newacronym{nssf}{NSSF}{Network Slice Selection Function}
\newacronym{o2i}{O2I}{Outdoor to Indoor}
\newacronym{oai}{OAI}{OpenAirInterface}
\newacronym{oaicn}{OAI-CN}{\gls{oai} \acrlong{cn}}
\newacronym{oairan}{OAI-RAN}{\acrlong{oai} \acrlong{ran}}
\newacronym{oam}{OAM}{Operations, Administration and Maintenance}
\newacronym{ofdm}{OFDM}{Orthogonal Frequency Division Multiplexing}
\newacronym{olia}{OLIA}{Opportunistic Linked Increase Algorithm}
\newacronym{omec}{OMEC}{Open Mobile Evolved Core}
\newacronym{onap}{ONAP}{Open Network Automation Platform}
\newacronym{onf}{ONF}{Open Networking Foundation}
\newacronym{onos}{ONOS}{Open Networking Operating System}
\newacronym{oom}{OOM}{\gls{onap} Operations Manager}
\newacronym{opnfv}{OPNFV}{Open Platform for \gls{nfv}}
\newacronym[type=hidden]{oran}{O-RAN}{Open \gls{ran}}
\newacronym{orbit}{ORBIT}{Open-Access Research Testbed for Next-Generation Wireless Networks}
\newacronym{os}{OS}{Operating System}
\newacronym{oss}{OSS}{Operations Support System}
\newacronym{pa}{PA}{Position-aware}
\newacronym{pase}{PASE}{Prioritization, Arbitration, and Self-adjusting Endpoints}
\newacronym{pawr}{PAWR}{Platforms for Advanced Wireless Research}
\newacronym{pbch}{PBCH}{Physical Broadcast Channel}
\newacronym{pcef}{PCEF}{Policy and Charging Enforcement Function}
\newacronym{pcfich}{PCFICH}{Physical Control Format Indicator Channel}
\newacronym{pcrf}{PCRF}{Policy and Charging Rules Function}
\newacronym{pdcch}{PDCCH}{Physical Downlink Control Channel}
\newacronym{pdcp}{PDCP}{Packet Data Convergence Protocol}
\newacronym{pdf}{PDF}{Probability Density Function}
\newacronym{pdsch}{PDSCH}{Physical Downlink Shared Channel}
\newacronym{pdu}{PDU}{Packet Data Unit}
\newacronym{pf}{PF}{Proportional Fair}
\newacronym{pgw}{PGW}{Packet Gateway}
\newacronym{phich}{PHICH}{Physical Hybrid ARQ Indicator Channel}
\newacronym{phy}{PHY}{Physical}
\newacronym{pmch}{PMCH}{Physical Multicast Channel}
\newacronym{pmi}{PMI}{Precoding Matrix Indicators}
\newacronym{powder}{POWDER}{Platform for Open Wireless Data-driven Experimental Research}
\newacronym{ppo}{PPO}{Proximal Policy Optimization}
\newacronym{ppp}{PPP}{Poisson Point Process}
\newacronym{prach}{PRACH}{Physical Random Access Channel}
\newacronym{prb}{PRB}{Physical Resource Block}
\newacronym{psnr}{PSNR}{Peak Signal to Noise Ratio}
\newacronym{pss}{PSS}{Primary Synchronization Signal}
\newacronym{pucch}{PUCCH}{Physical Uplink Control Channel}
\newacronym{pusch}{PUSCH}{Physical Uplink Shared Channel}
\newacronym{qam}{QAM}{Quadrature Amplitude Modulation}
\newacronym{qci}{QCI}{\gls{qos} Class Identifier}
\newacronym{qoe}{QoE}{Quality of Experience}
\newacronym{qos}{QoS}{Quality of Service}
\newacronym{quic}{QUIC}{Quick UDP Internet Connections}
\newacronym{rach}{RACH}{Random Access Channel}
\newacronym{ran}{RAN}{Radio Access Network}
\newacronym[firstplural=Radio Access Technologies (RATs)]{rat}{RAT}{Radio Access Technology}
\newacronym{rbg}{RBG}{Resource Block Group}
\newacronym{rcn}{RCN}{Research Coordination Network}
\newacronym{rc}{RC}{RAN Control}
\newacronym{rec}{REC}{Radio Edge Cloud}
\newacronym{red}{RED}{Random Early Detection}
\newacronym{renew}{RENEW}{Reconfigurable Eco-system for Next-generation End-to-end Wireless}
\newacronym{rf}{RF}{Radio Frequency}
\newacronym{rfc}{RFC}{Request for Comments}
\newacronym{rfr}{RFR}{Random Forest Regressor}
\newacronym{ric}{RIC}{RAN Intelligent Controller}
\newacronym{rlc}{RLC}{Radio Link Control}
\newacronym{rlf}{RLF}{Radio Link Failure}
\newacronym{rlnc}{RLNC}{Random Linear Network Coding}
\newacronym{rmr}{RMR}{RIC Message Router}
\newacronym{rmse}{RMSE}{Root Mean Squared Error}
\newacronym{rnis}{RNIS}{Radio Network Information Service}
\newacronym{rr}{RR}{Round Robin}
\newacronym{rrc}{RRC}{Radio Resource Control}
\newacronym{rrm}{RRM}{Radio Resource Management}
\newacronym{rru}{RRU}{Remote Radio Unit}
\newacronym{rs}{RS}{Remote Server}
\newacronym{rsrp}{RSRP}{Reference Signal Received Power}
\newacronym{rsrq}{RSRQ}{Reference Signal Received Quality}
\newacronym{rss}{RSS}{Received Signal Strength}
\newacronym{rssi}{RSSI}{Received Signal Strength Indicator}
\newacronym{rtt}{RTT}{Round Trip Time}
\newacronym{ru}{RU}{Radio Unit}
\newacronym{rw}{RW}{Receive Window}
\newacronym{rx}{RX}{Receiver}
\newacronym{s1ap}{S1AP}{S1 Application Protocol}
\newacronym{sa}{SA}{standalone}
\newacronym{sack}{SACK}{Selective Acknowledgment}
\newacronym{sap}{SAP}{Service Access Point}
\newacronym{sc2}{SC2}{Spectrum Collaboration Challenge}
\newacronym{scef}{SCEF}{Service Capability Exposure Function}
\newacronym{sch}{SCH}{Secondary Cell Handover}
\newacronym{scoot}{SCOOT}{Split Cycle Offset Optimization Technique}
\newacronym{sctp}{SCTP}{Stream Control Transmission Protocol}
\newacronym{sdap}{SDAP}{Service Data Adaptation Protocol}
\newacronym{sdk}{SDK}{Software Development Kit}
\newacronym{sdm}{SDM}{Space Division Multiplexing}
\newacronym{sdma}{SDMA}{Spatial Division Multiple Access}
\newacronym{sdn}{SDN}{Software-defined Networking}
\newacronym{sdr}{SDR}{Software-defined Radio}
\newacronym{seba}{SEBA}{SDN-Enabled Broadband Access}
\newacronym{sgsn}{SGSN}{Serving GPRS Support Node}
\newacronym{sgw}{SGW}{Service Gateway}
\newacronym{si}{SI}{Study Item}
\newacronym{sib}{SIB}{Secondary Information Block}
\newacronym{sinr}{SINR}{Signal to Interference plus Noise Ratio}
\newacronym{sip}{SIP}{Session Initiation Protocol}
\newacronym{siso}{SISO}{Single Input, Single Output}
\newacronym{sla}{SLA}{Service Level Agreement}
\newacronym{sm}{SM}{Service Model}
\newacronym{smf}{SMF}{Session Management Function}
\newacronym{smo}{SMO}{Service Management and Orchestration}
\newacronym{sms}{SMS}{Short Message Service}
\newacronym{smsgmsc}{SMS-GMSC}{\gls{sms}-Gateway}
\newacronym{snr}{SNR}{Signal-to-Noise-Ratio}
\newacronym{son}{SON}{Self-Organizing Network}
\newacronym{sptcp}{SPTCP}{Single Path TCP}
\newacronym{srb}{SRB}{Service Radio Bearer}
\newacronym{srn}{SRN}{Standard Radio Node}
\newacronym{srs}{SRS}{Sounding Reference Signal}
\newacronym{ss}{SS}{Synchronization Signal}
\newacronym{sss}{SSS}{Secondary Synchronization Signal}
\newacronym{st}{ST}{Spanning Tree}
\newacronym{svc}{SVC}{Scalable Video Coding}
\newacronym{tb}{TB}{Transport Block}
\newacronym{tcp}{TCP}{Transmission Control Protocol}
\newacronym{tdd}{TDD}{Time Division Duplexing}
\newacronym{tdm}{TDM}{Time Division Multiplexing}
\newacronym{tdma}{TDMA}{Time Division Multiple Access}
\newacronym{tfl}{TfL}{Transport for London}
\newacronym{tfrc}{TFRC}{TCP-Friendly Rate Control}
\newacronym{tft}{TFT}{Traffic Flow Template}
\newacronym{tgen}{TGEN}{Traffic Generator}
\newacronym{tip}{TIP}{Telecom Infra Project}
\newacronym{tm}{TM}{Transparent Mode}
\newacronym{to}{TO}{Telco Operator}
\newacronym{tr}{TR}{Technical Report}
\newacronym{trp}{TRP}{Transmitter Receiver Pair}
\newacronym{ts}{TS}{Technical Specification}
\newacronym{tti}{TTI}{Transmission Time Interval}
\newacronym{ttt}{TTT}{Time-to-Trigger}
\newacronym{tx}{TX}{Transmitter}
\newacronym{uas}{UAS}{Unmanned Aerial System}
\newacronym{uav}{UAV}{Unmanned Aerial Vehicle}
\newacronym{udm}{UDM}{Unified Data Management}
\newacronym{udp}{UDP}{User Datagram Protocol}
\newacronym{udr}{UDR}{Unified Data Repository}
\newacronym{ue}{UE}{User Equipment}
\newacronym{uhd}{UHD}{\gls{usrp} Hardware Driver}
\newacronym{ul}{UL}{Uplink}
\newacronym{um}{UM}{Unacknowledged Mode}
\newacronym{uml}{UML}{Unified Modeling Language}
\newacronym{upa}{UPA}{Uniform Planar Array}
\newacronym{upf}{UPF}{User Plane Function}
\newacronym{urllc}{URLLC}{Ultra Reliable and Low Latency Communications}
\newacronym{usa}{U.S.}{United States}
\newacronym{usim}{USIM}{Universal Subscriber Identity Module}
\newacronym{usrp}{USRP}{Universal Software Radio Peripheral}
\newacronym{utc}{UTC}{Urban Traffic Control}
\newacronym{vim}{VIM}{Virtualization Infrastructure Manager}
\newacronym{vm}{VM}{Virtual Machine}
\newacronym{vnf}{VNF}{Virtual Network Function}
\newacronym{volte}{VoLTE}{Voice over \gls{lte}}
\newacronym{voltha}{VOLTHA}{Virtual OLT HArdware Abstraction}
\newacronym{vr}{VR}{Virtual Reality}
\newacronym{vran}{vRAN}{Virtualized \gls{ran}}
\newacronym{vss}{VSS}{Video Streaming Server}
\newacronym{wbf}{WBF}{Wired Bias Function}
\newacronym{wf}{WF}{Waterfilling}
\newacronym{wg}{WG}{Working Group}
\newacronym{wlan}{WLAN}{Wireless Local Area Network}
\newacronym{osm}{OSM}{Open Source \gls{nfv} Management and Orchestration}
\newacronym{pnf}{PNF}{Physical Network Function}
\newacronym{drl}{DRL}{Deep Reinforcement Learning}
\newacronym{mtc}{MTC}{Machine-type Communications}
\newacronym{osc}{OSC}{O-RAN Software Community}
\newacronym{mns}{MnS}{Management Services}
\newacronym{ves}{VES}{\gls{vnf} Event Stream}
\newacronym{ei}{EI}{Enrichment Information}
\newacronym{fh}{FH}{Fronthaul}
\newacronym{fft}{FFT}{Fast Fourier Transform}
\newacronym{laa}{LAA}{Licensed-Assisted Access}
\newacronym{plfs}{PLFS}{Physical Layer Frequency Signals}
\newacronym{ptp}{PTP}{Precision Time Protocol}
\newacronym{cbrs}{CBRS}{Citizen Broadband Radio Service}
\newacronym{rnti}{RNTI}{Radio Network Temporary Identifier}
\newacronym{tbs}{TBS}{Transport Block Size}
\newacronym{ticc}{TICC}{Toeplitz Inverse Covariance-Based Clustering}

\newacronym{agv}{AGV}{Automated Guided Vehicle}
\newacronym{ar}{AR}{Augmented Reality}
\newacronym{gan}{GAN}{Generative Adversarial Network}
\newacronym{coreset}{CORESET}{Control Resource Set}
\newacronym{ks}{K-S}{Kolmogorov–Smirnov}
\newacronym{xai}{XAI}{EXplainable Artificial Intelligence}

\tikzstyle{startstop} = [rectangle, rounded corners, minimum width=2cm, minimum height=0.5cm,text centered, draw=black]
\tikzstyle{io} = [trapezium, trapezium left angle=70, trapezium right angle=110, minimum width=3cm, minimum height=1cm, text centered, draw=black]
\tikzstyle{process} = [rectangle, minimum width=2cm, minimum height=0.5cm, text centered, draw=black, alignb=center]
\tikzstyle{decision} = [ellipse, minimum width=2cm, minimum height=1cm, text centered, draw=black]
\tikzstyle{arrow} = [thick,<->,>=stealth]
\tikzstyle{line} = [thick,>=stealth]
\tikzstyle{darrow} = [thick,<->,>=stealth,dashed]
\tikzstyle{sarrow} = [thick,->,>=stealth]
\tikzstyle{larrow} = [line width=0.1mm,dashdotted,->,>=stealth]
\tikzstyle{llarrow} = [line width=0.1mm,->,>=stealth]

\makeatletter
\def\grd@save@target#1{%
  \def\grd@target{#1}}
\def\grd@save@start#1{%
  \def\grd@start{#1}}
\tikzset{
  grid with coordinates/.style={
    to path={%
      \pgfextra{%
        \edef\grd@@target{(\tikztotarget)}%
        \tikz@scan@one@point\grd@save@target\grd@@target\relax
        \edef\grd@@start{(\tikztostart)}%
        \tikz@scan@one@point\grd@save@start\grd@@start\relax
        \draw[minor help lines] (\tikztostart) grid (\tikztotarget);
        \draw[major help lines] (\tikztostart) grid (\tikztotarget);
        \grd@start
        \pgfmathsetmacro{\grd@xa}{\the\pgf@x/1cm}
        \pgfmathsetmacro{\grd@ya}{\the\pgf@y/1cm}
        \grd@target
        \pgfmathsetmacro{\grd@xb}{\the\pgf@x/1cm}
        \pgfmathsetmacro{\grd@yb}{\the\pgf@y/1cm}
        \pgfmathsetmacro{\grd@xc}{\grd@xa + \pgfkeysvalueof{/tikz/grid with coordinates/major step x}}
        \pgfmathsetmacro{\grd@yc}{\grd@ya + \pgfkeysvalueof{/tikz/grid with coordinates/major step y}}
        \foreach \x in {\grd@xa,\grd@xc,...,\grd@xb}
        \node[anchor=north] at (\x,\grd@ya) {\pgfmathprintnumber{\x}};
        \foreach \y in {\grd@ya,\grd@yc,...,\grd@yb}
        \node[anchor=east] at (\grd@xa,\y) {\pgfmathprintnumber{\y}};
      }
    }
  },
  minor help lines/.style={
    help lines,
    gray,
    line cap =round,
    xstep=\pgfkeysvalueof{/tikz/grid with coordinates/minor step x},
    ystep=\pgfkeysvalueof{/tikz/grid with coordinates/minor step y}
  },
  major help lines/.style={
    help lines,
    line cap =round,
    line width=\pgfkeysvalueof{/tikz/grid with coordinates/major line width},
    xstep=\pgfkeysvalueof{/tikz/grid with coordinates/major step x},
    ystep=\pgfkeysvalueof{/tikz/grid with coordinates/major step y}
  },
  grid with coordinates/.cd,
  minor step x/.initial=.5,
  minor step y/.initial=.2,
  major step x/.initial=1,
  major step y/.initial=1,
  major line width/.initial=1pt,
}
\makeatother

\acmConference[WiNTECH '24]{Proceedings of 18th ACM Workshop on Wireless Network Testbeds, Experimental Evaluation \& Characterization (WiNTECH '24)}{November 18, 2024}{Washington, D.C., USA}
\acmBooktitle{Proceedings of 18th ACM Workshop on Wireless Network Testbeds, Experimental Evaluation \& Characterization (WiNTECH '24), November 18, 2024, Washington, D.C., USA}

\ifexttikz
\else
\usepackage{tikzpagenodes,etoolbox}
\usetikzlibrary{calc}
\usepackage[contents={}]{background}
\AddEverypageHook{%
\ifnumequal{\thepage}{1}{%
    \tikz[remember picture,overlay]{%
        \node[draw,
        minimum width=1.03\textwidth,
        text width=1.02\textwidth,
        font=\footnotesize
        ]
        at ($(current page header area) - (0,-20pt)$)
        {%
        This paper has been accepted for publication on the Proceedings of the 18th ACM Workshop on Wireless Network Testbeds, Experimental evaluation \& Characterization (WiNTECH '24). This is the author's accepted version of the article. The final version published by ACM is L. Bonati, R. Shirkhani, C. Fiandrino, S. Maxenti, S. D'Oro, M. Polese, and T. Melodia, ``Twinning Commercial Network Traces on Experimental Open RAN Platforms'' \textit{WiNTECH '24: Proceedings of the 18th ACM Workshop on Wireless Network Testbeds, Experimental evaluation \& Characterization}, Washington, D.C., USA, November 2024.
        };
    }%
}{}
}
\fi

\begin{document}

\title{Twinning Commercial Network Traces on Experimental Open RAN Platforms}

\author[L. Bonati, R. Shirkhani, C. Fiandrino, S. Maxenti, S. D'Oro, M. Polese, T. Melodia]{Leonardo Bonati$^\dagger$,
Ravis Shirkhani$^\dagger$,
Claudio Fiandrino$^*$,
Stefano Maxenti$^\dagger$,
Salvatore D'Oro$^\dagger$,
Michele Polese$^\dagger$,
Tommaso Melodia$^\dagger$}
\affiliation{%
  \institution{$^\dagger$Institute for the Wireless Internet of Things, Northeastern University, Boston, MA, U.S.A.}
  \institution{$^*$IMDEA Networks Institute, Madrid, Spain}
  \city{}
  \state{}
  \country{}
  }

\acmYear{2024}\copyrightyear{2024}
\setcopyright{rightsretained}
\acmDOI{10.1145/3636534.3697320}
\acmISBN{979-8-4007-0489-5/24/11}

\begin{abstract}
While the availability of large datasets has been instrumental to advance fields like computer vision and natural language processing, this has not been the case in mobile networking. Indeed, mobile traffic data is often unavailable due to privacy or regulatory concerns.
This problem becomes especially relevant in Open \gls{ran}, where artificial intelligence can potentially drive optimization and control of the \gls{ran}, but still lags behind due to the lack of training datasets.
While substantial work has focused on developing testbeds that can accurately reflect production environments, the same level of effort has not been put into twinning the traffic that traverse such networks.

To fill this gap, in this paper, we design a methodology to twin real-world cellular traffic traces in experimental Open RAN testbeds.
We demonstrate our approach on the Colosseum Open \gls{ran} digital twin, and publicly release a large dataset (more than \datasethours and \datasetsize) with PHY-, MAC-, and App-layer \glspl{kpm}, and protocol stack logs.
Our analysis shows that our dataset can be used to develop and evaluate a number of Open \gls{ran} use cases, including those with strict latency requirements.
\end{abstract}

\ccsdesc[500]{Networks~Network measurement}
\ccsdesc[500]{Networks~Network performance analysis}
\ccsdesc[500]{Networks~Mobile networks}

\keywords{5G, Open RAN, Mobile Traffic Characterization, RAN Dataset.}

\settopmatter{printfolios=false}

\maketitle

\glsresetall
\glsunset{usrp}
\glsunset{fpga}
\glsunset{uhd}
\glsunset{gpu}

\section{Introduction}

The rise of big data, along with advancements in analytics and predictive modeling, has transformed research in computer vision, image processing, and Natural Language Processing (NLP) \cite{krizhevsky2017imagenet, devlin2018bert}. The availability of datasets at large~\cite{Kuznetsova_2020} led to significant progress in these fields, enabling also advancements of \gls{ai}/\gls{ml} techniques applied to these areas, thanks to the creation of common benchmarks fostering research reproducibility.

In mobile networks, the availability of datasets is more scarce as operators lack interest in publicly releasing such data for reasons that span from privacy concerns for the sensitivity of the data, to legal and regulatory aspects, as well as for strategic advantage over their competitors.
Nevertheless, initiatives that have made available mobile traffic data at metropolitan scale exist with a granularity of both multiple minutes (e.g., the Telecom Italia Big Data Challenge~\cite{tim-dataset} and NetMob~\cite{netmob23}) and milliseconds (e.g., the Madrid dataset~\cite{pablo-md}).
These datasets have been extremely helpful in advancing cellular technologies. For example, minute-level data can be used to optimize network deployment planning~\cite{francesco-tbd-netplan,gemmi2023globecom}, routing~\cite{commag-framework}, and to infer human and economic activities~\cite{zhang-tbd-urbdyn}. Similarly, millisecond-level data can be used to optimize resource allocation~\cite{nicola-tmc}, channel sounding~\cite{srs-allocation}, congestion control over mobile networks~\cite{xie-pbecc} or to understand specific mechanisms like network ID assignment to users~\cite{journal-giulia}.

The recent interest in Open \gls{ran}---and specifically O-RAN---deployments, where disaggregated cellular nodes can be reconfigured by \gls{ai}/\gls{ml} applications instantiated on \glspl{ric}, has resulted in the increased development of data-driven agents for \gls{ran} inference and control. However, the development, design, and training of these agents require massive amounts of data. Moreover, to ensure that agents can effectively adapt to a wide range of use cases and network conditions, such data needs to accurately reflect that of real-world deployments.

\noindent
\textbf{Related Work.}
Publicly available testbeds, such as Colosseum~\cite{villa2024dt} and the testbeds of the PAWR program~\cite{pawr}, in concert with open-source frameworks such as \openrangym~\cite{bonati2022openrangym-pawr}, provide reliable research platforms to collect data at scale in heterogeneous wireless deployments.
However, the quality of the collected data, and thus of the trained agents, is not solely related to the fidelity of the \gls{rf} setup, but also to that of the user traffic.
While a substantial body of work has focused on developing testbeds representative of real-world deployments~\cite{villa2024dt,breen2021powder,raychaudhuri2020cosmos,panicker2021aerpaw,zhang2021ara,villa2024x5g}, the generation of traffic that matches that of commercial networks is often overlooked.
Indeed, most traffic models do not reflect commercial traffic found in real-world deployments, or they do so in a coarse way, which is not suitable for experimentation on testbeds~\cite{meng2023modeling}.

Among the works that have tried to address this issue, a few leverage \glspl{gan} augmented with contextual information of the environment.
For instance, \cite{hui2023large} builds a knowledge graph to model spatial dependencies and content semantics, and uses it to generate cellular traffic.
\cite{zhang2023deep} proposes a deep transfer learning framework that uses historical information on existing cellular deployments to generate traffic for the planning of new sites by leveraging context information of source and target sites.
\cite{li2024mobile}~uses a \gls{lstm} network to capture the temporal correlation of traffic traces clustered from a real-world dataset, and then mimics their structure through a \gls{gan}.
However, these works do not focus on integrating the reproduced traces within simulators or platforms.

Similarly, \cite{meng2023modeling}~captures control-plane traffic of \glspl{ue} and models it via Semi-Markov models to evaluate and optimize core network deployments, such as SD-Core.
\cite{hsu2017hybrid}, instead, leverages datasets and traffic models to develop a \gls{mtc} traffic generator.
However, these works only focus on a single type of traffic.

Some works twin real-world traces into the ns-3 simulator.
For instance, \cite{lecci2021ns}~gathers data from virtual reality headsets and maps it into a burst traffic model.
\cite{agrawal2016trace} builds a framework to reproduce the characteristics of traces collected by users, while~\cite{5g-mmsys-europe} uses smartphones to generate and collect application traffic traces, such as file download and video streaming ones, and reproduce them in ns-3.
However, these works only focus on network simulators, with some of them considering traffic from a single application at a time. Instead, our approach is generic and accounts for traffic from the various concurrent applications that users are potentially running.

\noindent
\textbf{Contribution.}
In this paper, we bridge this gap by making a twofold contribution. First, we propose a methodology to twin real-world mobile traffic workloads in Open \gls{ran} platforms.
We do so by analyzing datasets gathered through cellular sniffers, and statistically reproducing the corresponding traffic through packet generation tools.
Then, we collect, analyze, and publicly release\footnote{The collected dataset is available at \url{https://github.com/wineslab/open-ran-commercial-traffic-twinning-dataset}.} a large dataset (more than \datasethours and \datasetsize of data) of cross-layer \gls{ran} \glspl{kpm} and protocol stack logs collected on an Open \gls{ran} deployment instantiated on Colosseum, where \glspl{ue} are served the twinned network traces. 
Specifically, we collect \bs- and \gls{ue}-level \glspl{kpm} from PHY, MAC, and App layers under different \gls{ran} configurations representative of \gls{ai}/\gls{ml} control policies, number of \glspl{ue}, and traffic demand.
Our dataset provides fine-grained and timestamped cross-layer metrics that make it possible to understand the connection between PHY and MAC \glspl{kpm} measured at the \bs and \glspl{ue}, control policies, and end-to-end and App-layer \glspl{kpm} that reflect user experience. Our analysis shows that PHY and MAC \glspl{kpm} alone are not representative of the user experience, and datasets also need to contain end-to-end \glspl{kpm} to properly capture the effect that control policies have on it.

\section{Primer on Dataset Collection Tools}

Passive LTE monitoring tools like FALCON~\cite{falcon} and LTESniffer~\cite{lte-sniffer-kaist} can be used to gather traffic traces from production \glspl{bs}. Decoding the \gls{pdcch}, which is sent without encryption, makes it possible to extract per-\gls{ue} scheduling information. While commercial tools such as Keysight WaveJudge exist, the landscape of open-source monitoring tools for 5G is yet to be shaped because of the complexity of decoding the 5G control channel, due to configuration flexibility and encryption of \gls{pdcch}. To the best of our knowledge, 5GSniffer~\cite{norbert-5gsniffer} is the only example of such monitoring tool, but it requires side-channel information about cell configuration in the event the production \gls{bs} varies the \gls{coreset} over time. 

In the case of FALCON, this tool runs on a Linux host connected to a \gls{sdr} and decodes the unencrypted LTE \gls{dci} at \gls{tti}-level. The procedure keeps the \gls{ue} identity anonymous and only shows its temporary ID (i.e., the \gls{rnti}). For each \gls{rnti}, the monitoring tool also provides the ID of the frame containing the traffic allocation, the associated \gls{tbs}, and transmission information, such as \gls{mcs} and utilized \glspl{prb}. As we will show, this information is sufficient to determine traffic characteristics at the \gls{bs}- and \gls{ue}-level, like the total traffic load or \gls{bs} utilization (at the \gls{bs} level) or the duration of per-user traffic bursts and idle times between subsequent traffic bursts (at the \gls{ue} level).

\section{From Real Traces to Open RAN Platforms}
\label{sec:from-real-traces-to-colosseum}

\begin{figure}[t]
\setlength\abovecaptionskip{4pt}
\setlength\belowcaptionskip{-5pt}
    \centering
    \includegraphics[width=\columnwidth]{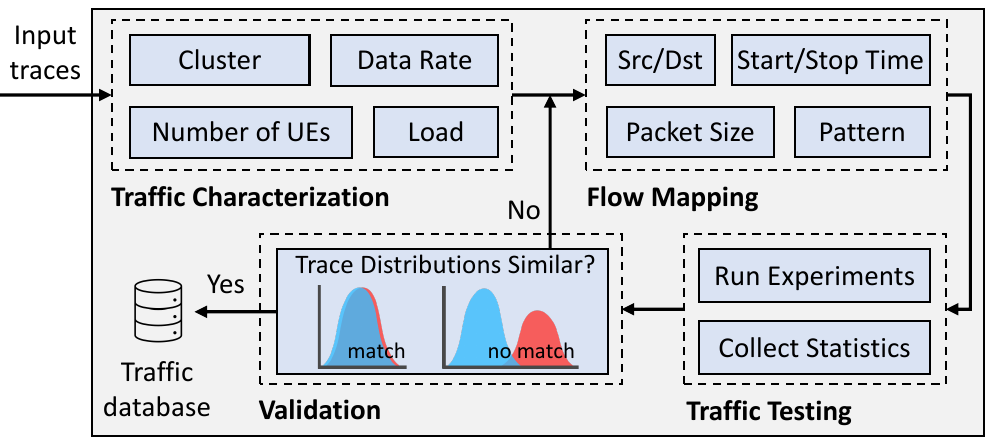}
    \caption{Pipeline to twin traffic traces from real-world datasets.}
    \label{fig:traffic-twinning-pipeline}
\end{figure}

Our pipeline to twin traces from real-world datasets is shown in Figure~\ref{fig:traffic-twinning-pipeline}.
At a high-level, it contains four main steps, described in the remaining of this section: traffic characterization, flow mapping, traffic testing, and validation. The goal of the pipeline is not to replicate transmissions on a packet-by-packet basis,\footnote{Note that this is generally not possible as traffic traces keep track of cell load and resource utilization only, but do not include transmitted packets.} but rather to identify traffic profiles that match those recorded in the wild and to statistically replicate them in Open \gls{ran} platforms.
Even though we focus on traces from an LTE commercial network, our procedure is generic and it can be applied to 5G traffic traces as well.

\subsection{Traffic Characterization}
\label{sec:characterizing-datasets}

We use a public dataset of LTE network obtained with FALCON from multiple \glspl{bs} located in different areas of Madrid, Spain~\cite{pablo-md}. The dataset contains the decoded control channel information of 6 different \glspl{bs}. In this paper, we focus on 3 \glspl{bs} from a suburban area of Madrid (BS$_1$, with carrier frequency at $816$~MHz, BS$_2$ at $1835$~MHz, and BS$_3$ at $2650$~MHz), which lets us cover three different spectrum bands.

\begin{figure}[t]
	\centering%
	\includegraphics[width=.95\columnwidth,keepaspectratio]{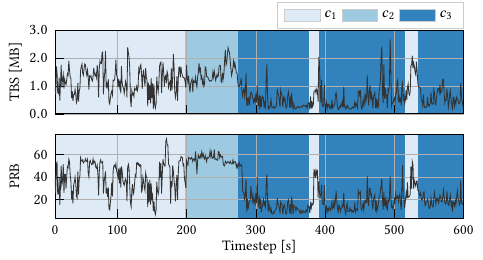}%
	\vspace*{-2ex}%
	\caption{Example of clustering of traffic of a production BS operating with a 20\:MHz channel bandwidth to identify slicing profiles.}%
	\label{fig:example-ticc}%
	\vspace*{-2ex}%
\end{figure}

At a high level, the traffic characterization follows four main steps: (i)~aggregate the raw data at the granularity of $1$\:s; (ii)~perform clustering on the pre-processed data (only needed if characterizing traffic profiles for network slices); (iii)~aggregate the pre-processed or clustered data over temporal windows $W$; and (iv) compute per-UE statistics suitable for flow mapping.
The input traces from the real-world dataset are loaded into memory by extracting a subset of data related to a temporal window of size $W$, which we treat as an aggregated data point to derive per-\gls{ue} statistics on the traffic to twin within the window.
We use the \gls{ticc} method to define traffic characteristics per slices, a key functionality of 5G networks~\cite{polese2021coloran} that, however, did not exist in LTE.
This technique segments multivariate time series into distinct clusters.
Rather than considering each point in isolation, \gls{ticc} uses a sliding window to group observations within their temporal context.
Since the original dataset comes with a too fine-grained resolution (i.e., at ms-level), we aggregate this information at $1$\:s granularity through a rolling-average strategy.\footnote{We experimented with much finer aggregation levels, e.g., at $10$\:ms, the duration of an LTE frame. In this case, \gls{ticc} takes longer to define the clusters, but we do not experience major differences in the cluster definition.}
As an example, Figure~\ref{fig:example-ticc} illustrates a trace for a \glspl{bs} of the dataset that can be divided in $C=3$ clusters with \gls{ticc}: $c_1$ is associated to conditions with high load and \gls{bs} resource utilization; $c_2$ is similar to $c_1$ but has a lower variability;  $c_3$ identifies periods where at least one variate is low.

\begin{figure}[ht]
\setlength\abovecaptionskip{4pt}
  \centering%
  \includegraphics[width=.95\columnwidth,keepaspectratio]{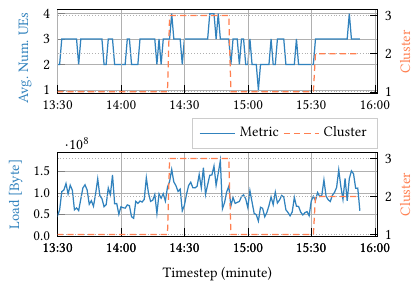}%
  \caption{Snapshot of traffic of a production BS. We report traffic load, average number of UEs and cluster for windows of $W=1$ minute.}%
  \label{fig:example-traffic-matching}%
\end{figure}

We derive information that includes the clusters that are present in the trace, data rate, load, and average number of \glspl{ue} for the considered window $W$. This procedure is necessary to identify the traffic profiles and their statistical properties, which are needed to twin the same statistical profile in the experimental platforms.  
An example is provided in Figure~\ref{fig:example-traffic-matching}, where we process data from the real traces of Figure~\ref{fig:example-ticc} using a window of size $W=1$\:minute. This makes it possible to convert the pre-processed data with granularity $1$\:s into aggregated per-\gls{ue} traffic profiles with larger granularity, and to capture their statistical properties. These can
be then used, for instance, to train \gls{ai}/\gls{ml} control policies.
Traces will be then used as input to the traffic generation tool on the experimental platforms, as discussed next.

\subsection{Flow Mapping}
\label{sec:flow-mapping}

Once the statistical behavior of each traffic flow is fully characterized, we map the statistical traffic flow information into traffic profiles compatible with \gls{mgen}~\cite{mgen_documentation}---a traffic generator developed by the U.S.\ Naval Research Laboratory---to generate data to be exchanged between \bs and \glspl{ue}. Source and destination of the real-world traces are mapped to IP addresses of the \bs and \glspl{ue} available in the testbed, together with the size of the packets and the start and stop time of each traffic flow expressed as the amount of seconds from the script start.
The traffic pattern also needs to be mapped to one supported by \gls{mgen}, which include Poisson-distributed, periodic, and burst traffic, among others.
Additional parameters, such as the number of parallel flows among \bs and \glspl{ue}, are also specified at this time.
Once this mapping is complete, \gls{mgen} scripts with instructions to generate the twinned traffic are built for every node used in the experiment performed on the testbed.

An example of an \gls{mgen} script that specifies per-\gls{ue} downlink traffic from the \bs is shown in Listing~\ref{lst:mgen-script}. For the sake of visualization, we only report instructions for the first $190$\:s of the experiment, while the complete script contains instructions for approximately $30$\:minutes.
\begin{lstlisting}[float=t,floatplacement=b,language=mgen,style=mystyle-mgen,
caption={Example of generated MGEN script to run at the BS. The script generates traffic flows for each UE connected to the BS.},
label={lst:mgen-script}]
70 ON 1 UDP DST 172.16.0.3/5000 PERIODIC [560.39 1250]
70 ON 2 UDP DST 172.16.0.4/5000 PERIODIC [560.39 1250]
70 ON 3 UDP DST 172.16.0.5/5000 PERIODIC [560.39 1250]
70 ON 5 UDP DST 172.16.0.7/5000 PERIODIC [10 125]
70 ON 6 UDP DST 172.16.0.8/5000 PERIODIC [10 125]
70 ON 7 UDP DST 172.16.0.9/5000 PERIODIC [10 125]

130 OFF 1
130 OFF 2
130 OFF 3
130 ON 1 UDP DST 172.16.0.3/5000 PERIODIC [512.05 1250]
130 ON 2 UDP DST 172.16.0.4/5000 PERIODIC [512.05 1250]
130 ON 3 UDP DST 172.16.0.5/5000 PERIODIC [512.05 1250]
130 ON 4 UDP DST 172.16.0.6/5000 PERIODIC [512.05 1250]
130 ON 8 UDP DST 172.16.0.10/5000 PERIODIC [10 125]

190 OFF 1
190 OFF 2
190 OFF 3
190 OFF 4
190 OFF 8
190 ON 1 UDP DST 172.16.0.3/5000 PERIODIC [555.73 1250]
190 ON 2 UDP DST 172.16.0.4/5000 PERIODIC [555.73 1250]
190 ON 3 UDP DST 172.16.0.5/5000 PERIODIC [555.73 1250]
\end{lstlisting}
In this example, we consider a scenario with a \bs and 8~\glspl{ue}, which corresponds to the Open \gls{ran} deployment used to test and validate the twinned traffic (see Section~\ref{sec:traffic-testing-validation}).
In accordance with the real-world traffic traces, the \glspl{ue} are divided in two service classes: \glspl{ue}~1-4 demand \gls{embb} traffic, \glspl{ue}~5-8 \gls{urllc} traffic.
Even though here we only illustrate how to map downlink traffic, it is worth noticing that the same approach can be applied to uplink traffic as well, and can be extended to additional traffic classes, \gls{rf} scenarios, and number of \glspl{ue}.
At second $70$ of the experiment, i.e., after allowing some time for the \glspl{ue} to connect to the \bs, traffic flows for \glspl{ue}~1-3 and~5-7 start. \gls{embb} \glspl{ue} are sent constant bitrate (called ``periodic'' by \gls{mgen}) traffic at $560.39$\:messages/s and messages of $1250$\:byte (\texttt{PERIODIC [560.39 1250]}), while \gls{urllc} \glspl{ue} are sent constant bitrate traffic at $10$\:messages/s and messages of $125$\:byte (\texttt{PERIODIC [10 125]}), as shown in lines~1-6. \glspl{ue}~4 and~8 are not sent any traffic initially, to be consistent with the real-world traces.
At second $130$, traffic flows for the \gls{embb} \glspl{ue} change. This is achieved by stopping the current flows (e.g., \texttt{130 OFF 1}) and starting new ones with different characteristics (\texttt{512.05 1250}). At the same time, two additional flows, one for an \gls{embb} \gls{ue} (\gls{ue}~4) and one for an \gls{urllc} one (\gls{ue}~8) are also started. This is shown in lines~8-15 of the listing. The flows for \glspl{ue}~5-7, instead, remain active.
Similarly, at second $190$, the flows for \glspl{ue}~1-3 are modified again, and those for \glspl{ue}~4 and~8 are stopped.
Overall, this lets us flexibly modify the traffic flows through scripted instructions that are then executed at run time. Further details on the \gls{mgen} syntax and capabilities can be found in the \gls{mgen} documentation~\cite{mgen_documentation}.

\subsection{Traffic Testing and Validation}
\label{sec:traffic-testing-validation}

After twinning the traces from the real-world dataset, we test the \gls{mgen} scripts of Section~\ref{sec:flow-mapping} by running experiments on the Open \gls{ran} platform.
We deploy a cellular network with \gls{bs} and \glspl{ue}, and leverage the \gls{mgen} scripts to generate traffic to be exchanged among them.
At the experiment run time, we collect traffic statistics and \glspl{kpm} from the protocol stack of the \gls{ran} nodes, and from \gls{mgen}.
These \glspl{kpm} correspond to those an operator would provide for \gls{ai}/\gls{ml} agent design, and that traffic sniffers can only partially capture.

Statistics and \glspl{kpm} are used to validate that the twinned traffic traces represent the profiles of the traces from the real-world dataset.
This is done by normalizing the twinned and real-world traffic distributions (e.g., their \glspl{pdf}) so that the x-axis takes values in $[0, 1]$, and comparing their similarity.
This can be done through methods such as the \gls{ks} test~\cite{massey1951kolmogorov} and by defining a similarity criterion (e.g., \gls{ks} distance below a user-defined threshold).
In case the similarity criterion is satisfied, the \gls{mgen} scripts are saved in a traffic database. Otherwise, the pipeline goes back to the flow mapping step of Section~\ref{sec:flow-mapping} to further tune the \gls{mgen} parameters, after which testing and validation are performed anew. This process repeats until the similarity criterion is satisfied.

\section{Colosseum Open RAN Dataset}
\label{sec:dataset}

We leverage the network traces twinned in Section~\ref{sec:from-real-traces-to-colosseum} to collect a large dataset of timestamped \glspl{kpm} from an Open \gls{ran} deployment instantiated on the Colosseum testbed. This dataset can be used to design and train \gls{ai}/\gls{ml} models to be run as O-RAN applications, e.g., xApps, rApps, and~dApps.

Colosseum is the largest Open \gls{ran} digital twin~\cite{villa2024dt}, with 128~pairs of compute nodes and \glspl{sdr} (USRP X310) interconnected through a channel emulator. Users can leverage softwarized frameworks to instantiate \bs and \gls{ue} protocol stacks, and control them through O-RAN applications deployed on the \glspl{ric}. Experimentation can be performed under a variety of \gls{rf} environments reproduced by a channel emulator, which is capable of emulating wireless channel effects such as path loss, multi-path, and fading. Different traffic profiles can also be emulated through a traffic emulation system based on \gls{mgen}, as well as via tools such as iPerf. 

We leverage \openrangym~\cite{bonati2022openrangym-pawr} to instantiate an Open \gls{ran} deployment on Colosseum and to perform an extensive data-collection campaign (more than \datasethours) to collect the dataset described in this section, which totals to more than \datasetsize of data.
For each experiment, we deploy a \bs and up to 8~\glspl{ue} belonging to the \gls{embb} and \gls{urllc} classes (4~\glspl{ue} each), which are allocated to separate slices of the \bs. \bs and \glspl{ue} are based on the srsRAN software, and are instantiated in an emulated \gls{rf} propagation environment corresponding to a cellular deployment in a neighborhood of Rome, Italy.
To be in line with the twinned real-world LTE dataset, our \bs leverages a frequency division duplexing configuration over $10$\:MHz of spectrum.
We use the \gls{mgen} scripts of Section~\ref{sec:flow-mapping} to generate downlink traffic flows among the \bs and \glspl{ue}.

We collect more than \datasetkpm timestamped \glspl{kpm} from the protocol stack of the \gls{bs} (reported by the \glspl{ue} or measured directly), from that of the \glspl{ue}, and from \gls{mgen}, under different number of \glspl{ue} and traffic demand.
Protocol stack \glspl{kpm} include PHY- and MAC-layer \glspl{kpm}, while \gls{mgen} ones are from the App layer. Our experiments span different clusters identified in the Madrid dataset, as well as different schedulers and resources allocated to the \gls{embb} and \gls{urllc} slices, which are representative of different \gls{ai}/\gls{ml} control policies.

\begin{table}[t]
\setlength\abovecaptionskip{2pt}
    \centering
    \footnotesize
    \setlength{\tabcolsep}{2pt}
    \caption{Resource allocation to the \acrshort{embb} and \acrshort{urllc} slices.}
    \label{tab:slicing-configurations}
    \begin{tabularx}{0.95\columnwidth}{
        >{\raggedright\arraybackslash\hsize=1.2\hsize}X
        >{\raggedleft\arraybackslash\hsize=0.9\hsize}X
        >{\raggedleft\arraybackslash\hsize=0.9\hsize}X }
        \toprule
        Slice Resource Allocation & \acrshort{embb} \acrshortpl{prb} & \acrshort{urllc} \acrshortpl{prb} \\
        \midrule
        \texttt{slicing\_1}     & 9     & 41 \\
        \texttt{slicing\_2}     & 21    & 29 \\
        \texttt{slicing\_3}     & 30    & 20 \\
        \texttt{slicing\_4}     & 39    & 11 \\
        \texttt{slicing\_5}     & 50    & 0 \\
        \bottomrule
    \end{tabularx}
\end{table}
Slice resources are computed in terms of \glspl{prb} that the \bs is allowed to use for each one of them out of a budget of 50~\glspl{prb} (i.e., $10$\:MHz of spectrum). The slicing configurations that we consider are shown in Table~\ref{tab:slicing-configurations}, while we consider \gls{rr} (scheduling~0) and \gls{pf} (scheduling~2) as scheduling algorithms.\footnote{\openrangym also includes the \gls{wf} algorithm (scheduling~1). Datasets including \gls{wf} are described in~\cite{bonati2021intelligence,polese2021coloran}, but they neither twin traffic from commercial traces, nor include App-layer KPMs.} In the dataset, \gls{embb} is marked as slice~0, \gls{urllc} as slice~1.
\glspl{ue} are allocated to the \gls{embb} and \gls{urllc} classes, and, hence, slices, based on their \gls{imsi}, as reported in Table~\ref{tab:user-slice-allocation}.
\begin{table}[t]
\setlength\abovecaptionskip{2pt}
    \centering
    \footnotesize
    \setlength{\tabcolsep}{2pt}
    \caption{\acrshort{ue} allocation to the \acrshort{embb} and \acrshort{urllc} classes of service.}
    \label{tab:user-slice-allocation}
    \begin{tabularx}{0.95\columnwidth}{
        >{\raggedright\arraybackslash\hsize=0.5\hsize}X
        >{\raggedright\arraybackslash\hsize=1.7\hsize}X
        >{\raggedleft\arraybackslash\hsize=0.9\hsize}X
        >{\raggedleft\arraybackslash\hsize=0.9\hsize}X }
        \toprule
        UE ID & UE IMSI & eMBB Class & URLLC Class \\
        \midrule
        1   & \texttt{1010123456002}     & x    & - \\
        2   & \texttt{1010123456003}     & x    & - \\
        3   & \texttt{1010123456004}     & x    & - \\
        4   & \texttt{1010123456005}     & x    & - \\
        5   & \texttt{1010123456006}     & -    & x \\
        6   & \texttt{1010123456007}     & -    & x \\
        7   & \texttt{1010123456008}     & -    & x \\
        8   & \texttt{1010123456009}     & -    & x \\
        \bottomrule
    \end{tabularx}
\end{table}
The most relevant PHY- and MAC-layer \glspl{kpm} collected from the protocol stacks of \bs and \glspl{ue},\footnote{BS KPMs are stored in files named \texttt{<ue\_imsi>\_metrics.csv} (one for each active UE) and in files named \texttt{enb\_metrics.csv} (cell-wide KPMs); UE KPMs in files named \texttt{ue\_metrics.csv}.} among which there are throughput, \gls{mcs}, and buffer occupancy, are reported in Table~\ref{tab:protocol-stack-metrics}.
\begin{table}[ht]
\setlength\abovecaptionskip{2pt}
    \centering
    \footnotesize
    \setlength{\tabcolsep}{2pt}
    \caption{Sample of per-UE protocol stack KPMs collected at the BS.}
    \label{tab:protocol-stack-metrics}
    \begin{tabularx}{\columnwidth}{
        >{\raggedright\arraybackslash\hsize=0.85\hsize}X
        >{\raggedright\arraybackslash\hsize=1.15\hsize}X }
        \toprule
        Metric & Description \\
        \midrule
        \texttt{dl\_buffer [bytes]}  & Occupancy in bytes of the downlink buffer queue with the data to be transmitted to the UE \\
        \texttt{dl\_mcs}     & Downlink MCS \\
        \texttt{tx\_brate downlink [Mbps]}   & Downlink throughput in Mbps  \\
        \texttt{tx\_pkts downlink}   & Number of downlink transmitted packets \\
        \texttt{dl\_cqi}   & Downlink CQI reported by the UE \\
        \texttt{sum\_requested\_prbs}   & Sum of the PRB needed to serve the \gls{ue}$^*$ \\
        \texttt{sum\_granted\_prbs} & Sum of the PRB granted to serve the \gls{ue}$^*$ \\
        \bottomrule
        \multicolumn{2}{>{\arraybackslash\hsize=2\hsize}X}{\scriptsize $^*$These are the total requested or granted PRBs over the $250$\:ms logging window.} \\
    \end{tabularx}
\end{table}
App-layer \glspl{kpm}, instead, are computed by the \gls{mgen} receiver running at each \gls{ue} that leverages information enclosed in the payload of the packets sent by the \gls{mgen} transmitter running at the \bs. These include the timestamp at which packets were sent and received, as well as sequence number and payload size (in bytes) for each packet, and allow for the computation of statistics such as latency, throughput and packet loss.\footnote{These metrics are recorded at each active UE in a file named \texttt{mgen.csv}.}

Finally, we also collect logs from the protocol stacks of \bs and \glspl{ue}. These are stored in files named \texttt{enb.log} and \texttt{ue.log}, respectively, and can be used, for instance, for the analysis of the \gls{rnti} timer expiration, as demonstrated in~\cite{journal-giulia}.

\section{Dataset Analysis}
\label{sec:results}

In this section, we overview relevant \glspl{kpm} collected in our dataset for the slicing configurations shown in Table~\ref{tab:slicing-configurations}. These metrics include downlink throughput for both MAC and App layers, \glspl{prb} allocated to the slices, end-to-end latency reported by \gls{mgen}, and \gls{cqi}.
While PHY- and MAC-layer \glspl{kpm} are reported directly in the dataset files, App-layer ones have been computed from the \gls{mgen} logs. We derived the end-to-end latency by computing the difference between the receive and transmit timestamp of each packet. For the throughput, instead, we computed the amount of data transmitted over windows of $250$\:ms, which have been selected to align the \gls{mgen} logs with the metrics reported by the protocol stacks of \bs and \glspl{ue}.

\begin{figure}[t]
\setlength\abovecaptionskip{1pt}
    \centering
    \includegraphics[width=\columnwidth,keepaspectratio]{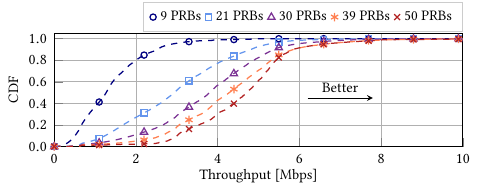}
    \caption{CDF of MAC-layer downlink throughput of eMBB UEs for different slicing configurations.}
    \label{fig:throughput_cdf_embb}
\end{figure}

Figure~\ref{fig:throughput_cdf_embb} shows the \gls{cdf} of the MAC-layer downlink throughput of the \gls{embb} \glspl{ue} for different slicing configurations.
We notice that larger \glspl{prb} allocations yield higher throughput values, since \glspl{ue} belonging to this class of service demand larger amounts of traffic.
Throughput of the \gls{urllc} slice, instead, is omitted as the traffic demand the \glspl{ue} is satisfied even with few \glspl{prb}.

\begin{figure}[ht]
\setlength\abovecaptionskip{1pt}
    \centering
    \includegraphics[width=\columnwidth,keepaspectratio]{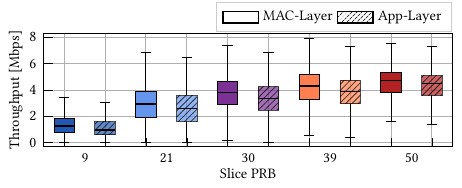}
    \caption{Bar plot of MAC- and App-layer downlink throughput of eMBB UEs for different slicing configurations.}
    \label{fig:throughput_box_embb_mac_app}
\end{figure}

Figure~\ref{fig:throughput_box_embb_mac_app} depicts the bar plot of the downlink throughput of the \gls{embb} \glspl{ue} at MAC (plain bars) and App (hatched bars) layers for the different slicing configurations of Table~\ref{tab:slicing-configurations}.
As expected, the throughput is higher at the MAC layer because of retransmissions to the \glspl{ue} in the downlink direction.

\begin{figure}[t]
\setlength\abovecaptionskip{1pt}
    \centering
    \includegraphics[width=\columnwidth,keepaspectratio]{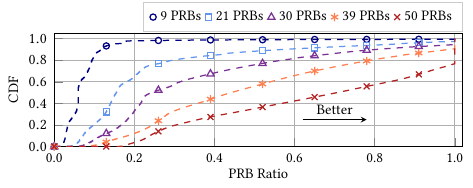}
    \caption{CDF of ratio between granted and requested PRBs for eMBB UEs for different slicing configurations.}
    \label{fig:prbs_ratio_embb}
\end{figure}

\begin{figure}[ht]
\setlength\abovecaptionskip{1pt}
    \centering
    \includegraphics[width=\columnwidth,keepaspectratio]{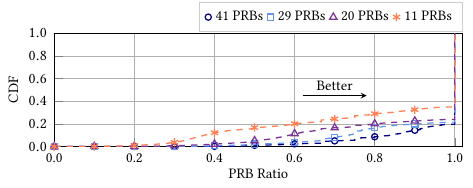}
    \caption{CDF of ratio between granted and requested PRBs for URLLC UEs for different slicing configurations.}
    \label{fig:prbs_ratio_urllc}
\end{figure}

We now investigate the ratio between the number of \glspl{prb} granted and requested by the \glspl{ue} (the higher, the better) for different slicing configurations.
\gls{cdf} results for the \gls{embb} \glspl{ue} are shown in Figure~\ref{fig:prbs_ratio_embb}, for the \gls{urllc} ones in Figure~\ref{fig:prbs_ratio_urllc}.
Notice that the \texttt{slicing\_5} configuration of Table~\ref{tab:slicing-configurations} is not shown for \gls{urllc}, as this would correspond to 0~\glspl{prb} allocated to this slice.
The \gls{prb} ratio gets higher as we allocate more \glspl{prb} to the \gls{embb} \glspl{ue} (Figure~\ref{fig:prbs_ratio_embb}). This means that with more resources available, we are more likely to satisfy the traffic demand of the \glspl{ue}.
This is consistent with the results of Figure~\ref{fig:throughput_cdf_embb}.
In the case of the \gls{urllc}, instead, fewer \glspl{prb} are enough to satisfy the traffic demand of \glspl{ue} (Figure~\ref{fig:prbs_ratio_urllc}), which are characterized by a less demanding traffic profile.

\begin{figure}[b]
\setlength\abovecaptionskip{1pt}
    \centering
    \includegraphics[width=\columnwidth,keepaspectratio]{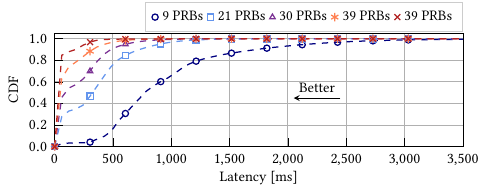}
    \caption{CDF of end-to-end latency of eMBB UEs for different slicing configurations.}
    \label{fig:latency_embb}
\end{figure}

\begin{figure}[t]
\setlength\abovecaptionskip{-2pt}
\setlength\belowcaptionskip{-5pt}
    \centering
    \includegraphics[width=\columnwidth,keepaspectratio]{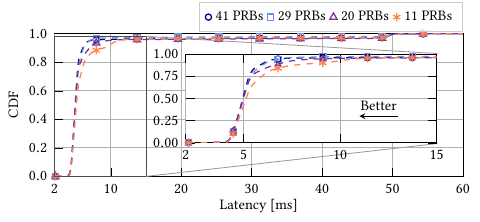}
    \caption{CDF of end-to-end latency of URLLC UEs for different slicing configurations.}
    \label{fig:latency_urllc}
\end{figure}

The \gls{cdf} of the end-to-end latency between the \bs and \glspl{ue} is shown in Figures~\ref{fig:latency_embb} (for \gls{embb}) and~\ref{fig:latency_urllc} (for \gls{urllc}).
Similarly to what happens in Figure~\ref{fig:prbs_ratio_urllc}, the \texttt{slicing\_5} case is not shown for the \gls{urllc} case.
The latency decreases when more \glspl{prb} are allocated to the slices, with the worst-case latency being within $2$\:s for \gls{embb} ($9$~\glspl{prb} case), and $10$\:ms for \gls{urllc} \glspl{ue} ($11$~\glspl{prb} case) with probability~$0.93$.

Table~\ref{tab:urllc-use-cases} shows the probability that the \gls{urllc} \glspl{ue} of our dataset meet the latency requirements of relevant 5G use cases defined in~\cite{ngmn2019verticals} for the different slicing configurations of Table~\ref{tab:slicing-configurations}.
\begin{table}[t]
\setlength\abovecaptionskip{2pt}
    \centering
    \footnotesize
    \setlength{\tabcolsep}{2pt}
    \caption{Probability of URLLC UEs satisfying end-to-end latency requirements (Req.) for different use cases in~\cite{ngmn2019verticals}.}
    \label{tab:urllc-use-cases}
    \begin{tabularx}{\columnwidth}{
        >{\raggedright\arraybackslash\hsize=2.45\hsize}X
        >{\raggedleft\arraybackslash\hsize=0.71\hsize}X
        >{\raggedleft\arraybackslash\hsize=0.71\hsize}X
        >{\raggedleft\arraybackslash\hsize=0.71\hsize}X
        >{\raggedleft\arraybackslash\hsize=0.71\hsize}X
        >{\raggedleft\arraybackslash\hsize=0.71\hsize}X }
        \toprule
        Use Case & Req. &  41~PRBs & 29~PRBs  & 20~PRBs & 11~PRBs \\
        \midrule
        AGV control & $5$\:ms & $0.625$ & $0.613$ & $0.474$ & $0.598$ \\
        Cloud gaming & $7$\:ms & $0.957$ & $0.929$ & $0.931$  & $0.887$ \\
        Robot tooling & $10$\:ms & $0.960$ & $0.971$ & $0.955$ & $0.948$ \\
        AR in smart factory & $15$\:ms & $0.964$ & $0.974$ & $0.959$  & $0.961$ \\
        Fault mgmt in distributed power generation & $30$\:ms & $0.968$ & $0.977$ & $0.963$ & $0.970$ \\
        UAV command and control & $100$\:ms & $0.999$ & $0.999$ & $0.999$ & $0.999$ \\
        Fault location identification & $140$\:ms & $0.999$ & $0.999$ & $0.999$ & $0.999$ \\
        \bottomrule
        \end{tabularx}
    \vspace{-12pt}
\end{table}
Use cases span from \gls{agv} control with a $5$\:ms end-to-end latency requirement, to cloud gaming ($7$\:ms), robot tooling ($10$\:ms), \gls{ar} for smart factories ($15$\:ms), automated fault management for power distribution grids ($30$\:ms), \gls{uav} command and control ($100$\:ms), and identification of fault location along electricity lines ($140$\:ms).
We notice that our dataset meets the requirements of most of the \gls{urllc} verticals shown in the table, with the 41~and 29~\glspl{prb} configurations (\texttt{slicing\_1} and \texttt{slicing\_2} in Table~\ref{tab:slicing-configurations}) outperforming the other ones. In general, more \glspl{prb} improve satisfaction of stringent latency requirements (e.g., \gls{agv} control, gaming), while less stringent latency requirements can be satisfied even with few \glspl{prb} allocated to \gls{urllc} \glspl{ue} (e.g., 11 \glspl{prb} in the case of latency requirements $\ge 100$\:ms).

\begin{figure}[t]
\setlength\abovecaptionskip{1pt}
\setlength\belowcaptionskip{-15pt}
    \centering
    \includegraphics[width=\columnwidth,keepaspectratio]{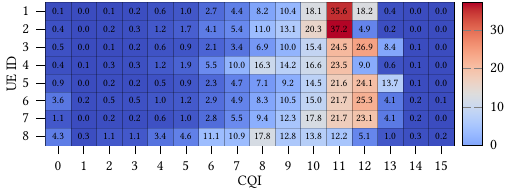}
    \caption{Percentage of times UEs report certain CQI values. CQI~0 indicates that the UE is out of range. UEs~1 and~2 are closer to the BS.}
    \label{fig:cqi_heatmap}
\end{figure}

Finally, Figure~\ref{fig:cqi_heatmap} shows the heat map of the percentage of times the \glspl{ue} report a certain \gls{cqi} value to the \gls{bs}, aggregated over all the slicing configurations of Table~\ref{tab:slicing-configurations}. 
The \gls{cqi} is reported as an integer number from~0 to~15 (the higher, the better), each associated to a specific modulation that will be used for downlink transmissions.
We notice that in most of the cases, \glspl{ue} report a \gls{cqi} between~8 and~12.

\vspace{-5pt}
\section{Discussion}

Although we focused on an LTE dataset~\cite{pablo-md}, our methodology is general.
However, starting from an LTE dataset did not allow us to capture some effects typical of 5G networks, such as multiple threads with distinct traffic patterns.
In future works, we plan to apply our approach to datasets collected with 5G sniffers~\cite{norbert-5gsniffer} and integrate them in Colosseum.

This work offers a tool to replicate data collected over the air in commercial deployments on Open \gls{ran} digital twins, and the possibility to convert static traces into dynamic traffic scenarios that can be used for a variety of tasks. 
For example, datasets generated with our methodology can be used as a data-augmentation mechanism for improving \gls{ai}/\gls{ml}-based forecasting and classification.
Similarly, our methodology and dataset can also help in advancing \gls{xai}, as the twinned traffic can be used to identify small- and macro-scale events, properties, and transitions that cause accuracy drop or unexpected behavior. 

Finally, our dataset can be used to generate novel \gls{ai}/\gls{ml} algorithms for purposes that go beyond forecasting and traffic analytics, e.g., O-RAN-enabled control. Indeed, static traces are useful to train forecasters and classifiers, but are not well-suited to assess the effectiveness of control policies, and their impact on network performance due to their static nature. Our dataset, instead, has been generated by considering a large number of control policy configurations, which can truly offer insights on which control action delivers the best performance under varying network and traffic conditions.

\vspace{-5pt}
\section{Conclusions}

In this paper, we proposed a methodology to twin commercial traffic traces from real-world datasets into Open \gls{ran} testbeds to improve the accuracy of cellular experimental research.
We demonstrated our approach on the Colosseum Open \gls{ran} digital twin, where we collected a large dataset with cross-layer \glspl{kpm} that we publicly released.
Our dataset can be used for the design and evaluation of Open \gls{ran} use cases, including those with strict latency requirements.

\vspace{-5pt}
\begin{acks}
This work was partially supported by the O-RAN ALLIANCE, by the U.S.\ National Science Foundation under grants CNS-1925601, CNS-2112471 and CNS-2120447, by the U.S.\ National Telecommunications and Information Administration (NTIA)'s Public Wireless Supply Chain Innovation Fund (PWSCIF) under Award No. 25-60-IF011, by the bRAIN project PID2021-128250NB-I00 funded by \url{MCIN/AEI/10.13039/501100011033}, and by the European Union ERDF ``A way of making Europe.'' C. Fiandrino is a Ramón y Cajal awardee (RYC2022-036375-I), funded by \url{MCIU/AEI/10.13039/501100011033} and the ESF+.
\end{acks}

\vspace{-5pt}
\balance
\bibliographystyle{ACM-Reference-Format-shortened}  
\bibliography{biblio}

\end{document}